\documentstyle[preprint,eqsecnum,aps,overcite]{revtex}
\begin{document}
\title{
COSMIC CONCORDANCE
}
\author{J. P. Ostriker}
\address{
Department of Astrophysical Sciences \\
 Princeton University  \\
Princeton, N.J. 08544 USA
}
\author{Paul J. Steinhardt}
\address{
Department of Physics and Astronomy\\
University of Pennsylvania\\
Philadelphia, Pennsylvania 19104  USA
}

\maketitle
\begin{abstract}
{\it It is interesting, and perhaps surprising, that despite a
growing diversity of independent astronomical and
cosmological observations, there remains a
substantial range of cosmological models  consistent with all important
observational constraints.  The constraints guide one forcefully to
examine models in which the matter density is substantially less than critical
density.
Particularly noteworthy are those which are consistent with inflation.
For these models, microwave background  anisotropy, large-scale structure
measurements,
direct measurements of the Hubble constant,
$H_0$, and the closure parameter, $\Omega_{\rm Matter}$, ages of
 stars and a host of more minor facts are all consistent
with a spatially flat model having significant cosmological constant
$\Omega_{\Lambda} = 0.65 \pm 0.1$, $\Omega_{\rm Matter} = 1 - \Omega_{\Lambda}$
(in the form of ``cold dark matter") and a small tilt:  $0.8< n  < 1.2$.}
\end{abstract}

\newpage

	Our approach has been to objectively apply the best known
observational constraints to classes of theoretical models and
map out any permitted range of parameters.  In addition to the
astronomical measurements that have been standardly considered,
we incorporate recent measurements of the cosmic background
radiation (CBR) anisotropy at large, intermediate and small
angular scales.
We present the results of  a most promising case, spatially flat
models with cold dark matter and cosmological constant.
These models are consistent with a primary prediction of the
inflationary paradigm\cite{1} (for a recent review, see Ref. 2); that is, to
high precision, $\Omega_{\Lambda} + \Omega_{\rm m}+\Omega_{\rm rad} = 1$
 today.   Although the common theoretical prejudice is
that $\Omega_{\Lambda}=0$,
 we consider it plausible that there is a new energy
scale (few meV) being unveiled, associated with
weakly coupled fields ({\it e.g.} neutrinos or dark matter).

	We  first consider astronomical measurements that
delimit the ($\Omega_{0,{\rm Matter}}$,~$h$) plane
(Fig. 1).   For the Hubble parameter, most recent
observations\cite{3} are in the
range $h =0.70 \pm 0.15$, including the major recent study
using the Hubble Space Telescope and the classical
Cepheid variables\cite{4} ($h= 0.82 \pm 0.17$)
and studies using Type I supernovae\cite{41} ($h =  0.67 \pm 0.07$).
We will take as a lower bound on the age of the universe the
ages of the oldest globular clusters:  $t_0 = 15.8 \pm 2.1$ based on
the main sequence turnoff\cite{5}
 and $t_0 = 13.5 \pm  2.0$ using giant branch fitting.\cite{6}
(The 11.5 Gyr lower bound is illustrated in Fig. 1; using the
13.7 Gyr value would reduce the concordance region by roughly
half, still maintaining a substantial area.)

	The cosmological constant,  $\Omega_{\Lambda}$,
is constrained by many tests,\cite{7}
but most directly by gravitational lensing which
gives $\Omega_{0,\Lambda} < 0.75$.\cite{8}
This is consistent with the lower bound  $0.2  <
\Omega_{0,{\rm Matter}}$  based on observed light density and
cluster mass-to-light ratios\cite{10}
or by utilization of large-scale structure measurements.\cite{11}

	  The X-ray measured gas masses and the total
virial masses\cite{12} imply
$\Omega_{B} h^{3/2}/\Omega_{0,{\rm Matter}}  = 0.07 \pm 0.03$,
and light element nucleosynthesis\cite{121}
constrains $\Omega_{B} h^2  = 0.015 \pm 0.005$,
where $\Omega_{B}$ is the baryonic contribution to the critical density.
  These combine into a constraint on
$1- \Omega_{0,\Lambda} = \Omega_{0,{\rm Matter}} =
(0.21 \pm 0.12)h^{-1/2}$.
Finally, the growth of large scale structure in CDM models\cite{13}
requires $\Gamma \equiv \Omega_{0,{\rm Matter}} h = 0.25 \pm 0.05$.

\begin{figure}
\caption{
The range of models (hatched area) in concordance with the
best known astronomical observations of the Hubble constant ($H_0$),
age (allowed region is {\it above} the dashed curve), large scale
structure ($\Gamma \equiv \Omega_{\rm Matter}h$), baryons in galaxy
clusters (dot-dashed curves), and gravitational lenses (allowed models
are below the shaded region).  Dot indicates a representative model
with $h=0.65$ and $\Omega_{\Lambda}=0.65$.}
\end{figure}

	 The range of concordance with all the quoted
observations is indicated by  the hatched area in Fig. 1.
The black dot denotes a representative model
$h = 0.65$, $\Omega_{\Lambda} = 0.65$.
Two conclusions are worth noting here:
 (1) a substantial permitted area does exist, and
 (2) removal of any one of the observational sets of
limits does not significantly enlarge the
 permitted area.  Or, put differently,
 recovering the more theoretically desirable
$\Omega_{\Lambda} = 0$
requires a combination of many observations
to change in a coherent fashion.  Using similar astronomical
arguments, others have been led to non-zero
$\Omega_{\Lambda}$ models independently.\cite{14}

	We now add the cosmological constraints derived from  CBR
anisotropy.     The amplitude of the CBR power spectrum can
be used to determine the value of $\sigma_8$,
the fluctuations in mass in an eight $h^{-1}$~Mpc sphere.
The extrapolation from the large angular scales probed
by COBE DMR down to 8 Mpc depends on the value of the spectral index
$n$, the fractional contribution of gravitational
waves to CBR fluctuations, and the values of
$\Lambda$, $h$,  and $\Omega_{B}$.\cite{17}
   For any given point in the concordance region of Fig. 1, an
allowed range for $n$
can be determined by extrapolating  the COBE DMR amplitude down to
8 Mpc and comparing to the range
$\sigma_8 = (0.56 \pm 0.06) (\Omega_{0,{\rm Matter}})^{-0.56}$
 derived from the great clusters and from the
distribution of large-scale structure and velocities.\cite{11}
 Here we explicitly assume inflation, which fixes a relation
between $n$  and the gravitational wave contribution.\cite{2}$^,$\cite{19}
 We find agreement for tilts,
$- 0.2 < n - 1 < 0.2$, consistent with COBE DMR spectral index
measurements\cite{18}
and with what is achievable in inflationary models.\cite{2}$^,$\cite{19}
Hence, we find that cosmic concordance with all observations
and with inflationary cosmology can be obtained.

	We might have taken as a postulate that
$\Omega_{\Lambda} = 0$  and considered open universe models
instead.  One problem is that open universes can be
accommodated  with inflation only by very delicate tuning of parameters.
Also, the diagram analogous to Figure 1, would show a much
smaller concordance region because the age constraint would have
shifted substantially to the left.   We do not intend to rule out the
possibility  that yet other types of models, such as mixed
dark matter or baryonic isocurvature, might be constructed to
fit the observational constraints.

	What future observations could be used to further
reduce the concordance range?  The most promising are  measurements
of intermediate- and small-scale CBR anisotropy.  Figure 2
shows the predicted CBR anisotropy power spectrum\cite{20}
 for the standard CDM model  and a representative model from the
middle of the concordance region
$(h=0.65, \Omega_{\Lambda}=0.65)$.
A characteristic feature of inflationary models and a spatially flat universe
is the Doppler peak at $\ell \approx 220$.
Hence, simply finding (or not finding) a Doppler peak at $\ell \approx 220$
 would rule out either all open (or the flat) models.

\begin{figure}
\caption{
 Predicted CBR power spectrum showing the spectrum of
multipoles (${\cal C}_{\ell}$) as  a function of multipole
number ($\ell$) for standard CDM ($n=1$, $h=0.5$, $\Omega_{\Lambda}=0$;
dashed line) and a representative concordance model ($n=0.96$, 20\%
gravity wave contribution to CBR quadrupole, $h=0.65$,
$\Omega_{\Lambda}=0.65$, $\sigma_8=0.87$; solid line).
The boxes represent the theoretical predictions for present CBR
experiments. The horizontal error bars are present one-sigma detections,
and the triangles are 95 percent upper confidence limits.  (See
Ref. 2 for details.) Note that the power spectrum for the concordance
model is remarkably similar to the prediction for CDM, except at
smaller angular scales ($\ell>250$), where the concordance model is
marginally more consistent with present observational limits.
}
\end{figure}

	Experiments at yet smaller angular scales,
down to 10 arcminutes, would provide further
constraints on cosmological parameters.\cite{2}$^,$\cite{20}
  An amusing feature, illustrated in Figure 2, is that the predicted
CBR power spectrum for
$\ell < 250$
in standard ($\Omega_{\Lambda}=0$) CDM models and in our concordance models
are virtually indistinguishable and in equal agreement
with observations, despite their substantially different
cosmological parameter values.  However, for
$\ell > 250$ (spanning $10'$-$30'$),
representative concordance models predict somewhat less power than standard
CDM,  marginally more consistent with current measurements.\cite{2}
This example is  strong motivation  for improved CBR
experiments with $10'$-$30'$ angular resolution.

	We have presented this analysis, in part,
to lay down a challenge to the reader:
Can one identify a serious problem with the concordance models illustrated
in Figs. 1 and 2?  If not, perhaps we have already identified
models which, in broad outline, capture the essential
properties of the large-scale universe.
More generally, we offer this analysis as a forward-looking
illustration of  how new and improved observations will
provide quantitative and redundant tests that can
decisively discriminate among competing models.

\noindent
ACKNOWLEDGEMENTS:
	The work of P. J. Steinhardt was supported by the
John S. Guggenheim Foundation and  the National Science
Foundation
at the Institute for Theoretical Physics (ITP) at
Santa Barbara and at the Institute for Advanced Studies
at Princeton.  The work of J. P. Ostriker was supported by
NASA and the National Science Foundation.
We  wish to thank R. Bond, R. Cen, P.J.E. Peebles, E.
Turner, S. White and E. Witten for their
useful conversations, and also the
ITP which fostered the collaboration.
\end{document}